# Comment on "Bubble nucleation and cooperativity in DNA melting"

# [Phys. Rev. Letters 94, 035504 (2005)]

Michaël SANREY and Marc JOYEUX[(#)]

*Laboratoire de Spectrométrie Physique, Université Joseph Fourier,*

*38402 St Martin d'Hères, France*

**Abstract** : The conclusions presented in this Letter rely on not converged calculations and should be considered with caution.

Zeng *et al* investigated the temperature evolution of the length of bubbles close to the melting of short DNA sequences and concluded that the average size of "bubbles in the middle" reaches a plateau equal to the number of AT pairs before the two strands separate [1]. Ares *et al* performed Monte Carlo (MC) simulations with the Dauxois-Peyrard-Bishop (DPB) model and claimed that they support experimental observations [2]. In contrast, Van Erp *et al* showed that intermediate states are not observed when the DPB model is investigated with the Transfer Integral (TI) method and tentatively ascribed this difference to their use of (i) a slightly different definition of the fractional bubble length $\langle \ell \rangle$, (ii) open instead of periodic boundaries, (iii) a bias potential [3]. In this Comment we point out that the difference more likely arises from the lack of convergence of the results in [2].

We got results comparable to [2] when performing a moderate number of Metropolis steps ($\approx 10^6$) but curves obtained with $5.10^8$ steps for both thermal equilibration and averaging, and maximum random displacements of $\mp 1.5$ Å, are different [4]. For example, *f*, *p* and $\langle \ell \rangle$ (estimated as in [2]) are shown in Fig. 1 for the L42B18 sequence [1] with periodic boundaries. Compared to Fig. 1 of [2], these curves are steeper and shifted by 20°C to lower temperatures, which reflects the fact that an increasing number of sequences separate further and further, thus letting the partition function diverge. Moreover, *f* and *p* remain very close ($f - p < 0.03$) and $\langle \ell \rangle$ cannot be calculated in the interval where the plateau is expected because $f \approx p \approx 1$. The curves in Fig. 1 therefore do not support the onset of intermediate states and the differences with [2] become even more pronounced for larger numbers of steps. Divergence of the partition function can be forbidden by imposing an upper value for strand separation, but we checked that a threshold of 50 Å leads to curves very close to Fig. 1.

The lack of convergence in [2] is due to the fact that the authors used Eq. 7 of [5] instead of the standard formula for new positions. Replacing $y_i$ by $(y_{i-1} + y_{i+1})/2$ amounts to introducing a barrier to dissociation, so that sampling with the modified formula leads to a distribution that differs a lot from Bolzmann's one. Moreover, results depend critically on the assumed values for the threshold and the width of the Gaussian distribution.

On the other hand, converged values of $\langle \ell \rangle$ can be obtained by averaging the fraction of open base pairs over the MC steps where the separation of at least one base pair is smaller than the dissociation threshold, which is equivalent to working in the "double-stranded DNA ensemble" introduced in [3]. Results obtained with this method (Fig. 2) are in excellent agreement with the TI calculations reported in Fig. 4 of [3]. These curves do not display any plateau and therefore do not support the onset of intermediate states.

# REFERENCES

[1] Y. Zeng *et al*, Phys. Rev. Lett. 91, 148101 (2003)

[2] S. Ares *et al*, Phys. Rev. Lett. 94, 035504 (2005)

[3] T.S. van Erp *et al*, Eur. Phys. J. E 20, 421 (2006)

[4] See EPAPS Document No. [XXX] for the Pascal code used to obtain the results in Fig. 1.

[5] S. Ares *et al*, Phys. Rev. E 67, 046108 (2003)

# FIGURE CAPTIONS

**Figure 1** : $f$, $p$ and $\langle \ell \rangle$ as a function of $T$ for the L42B18 sequence with periodic boundaries and dissociation thresholds of 0.5 and 2.0 Å.

**Figure 2** : $\langle \ell \rangle$ as a function of $T$, for the L42B18 sequence with open and periodic boundaries and dissociation thresholds of 0.5 and 2.0 Å.

**FIGURE 1**

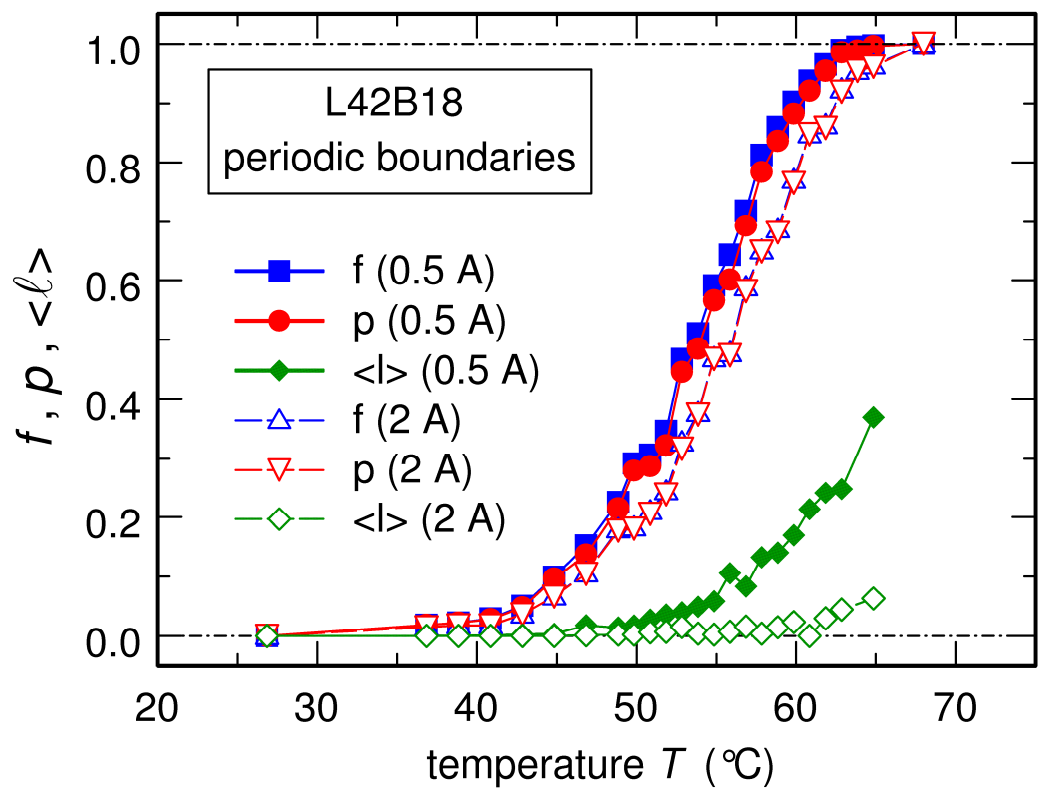

**FIGURE 2**

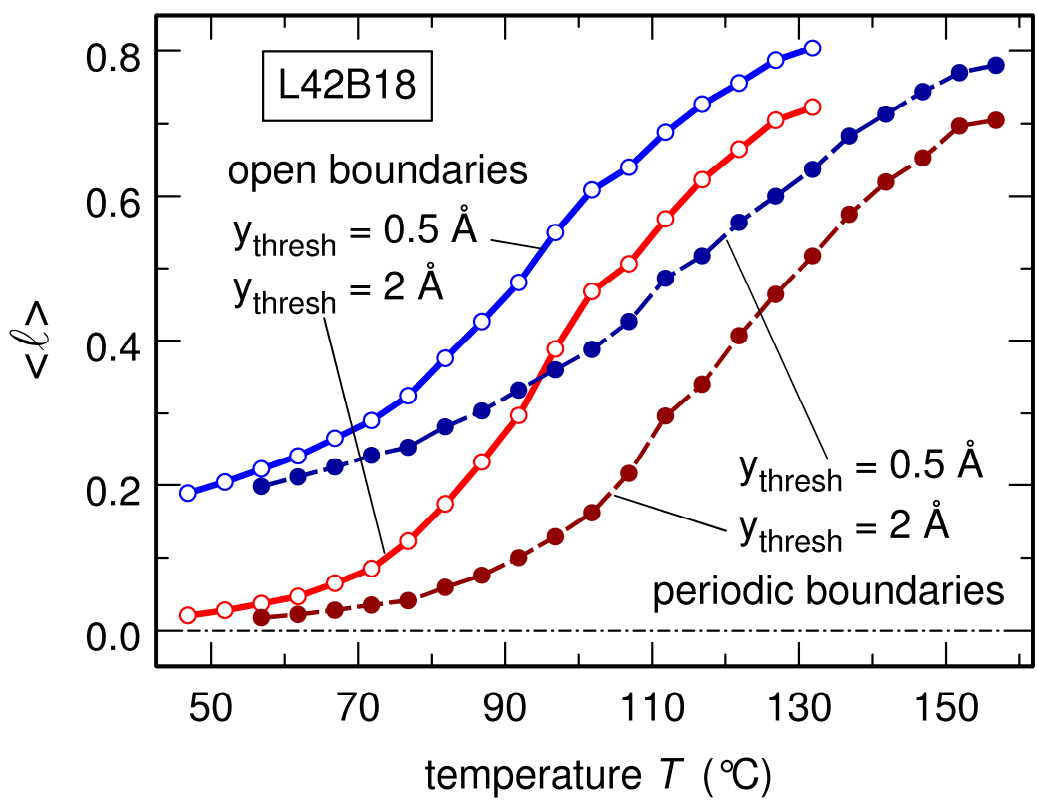